\documentclass[letter,traditabstract]{aa}
\usepackage{txfonts}
\usepackage{graphicx}
\usepackage{natbib}
\bibpunct{(}{)}{;}{a}{}{,} 
\usepackage{epsfig}

\begin{document}
\title{The Time Dependence of hot Jupiters' Orbital Inclinations
}
\author{Amaury H.\,M.\,J. Triaud
}
\offprints{Amaury.Triaud@unige.ch}

\institute{Observatoire Astronomique de l'Universit\'e de Gen\`eve, Chemin des Maillettes 51, CH-1290 Sauverny, Switzerland
}

\date{Received date / accepted date}
\authorrunning{Triaud A.\,H.\,M.\,J.}
\titlerunning{Age and spin/orbit angle}

\abstract{
Via the Rossiter-McLaughlin effect, it is possible to measure the sky-projected  angle between the stellar spin and a planet's orbital spin. Observed orbital inclinations have been found to range over all possible angles. 
A tentative detection of a correlation between the dispersion in spin/orbit angle and the youth of the system is revealed, using spin/orbit measurements for hot Jupiters around stars with masses $\geq 1.2\,M_\odot$ for which age estimates are more accurately determined. The chance of this pattern arising by chance has been computed to 7\%. 
This appears in accordance with tidal dissipation where non-coplanar hot Jupiters' orbits tidally realign. The results show they would do so within about 2.5 Gyr. 
For the considered sample, the results give support to hot Jupiters being placed on non coplanar orbits early in their history rather than this happening late. 
Such events could involve strong planet-planet scattering. 

 }

\maketitle

\section{Introduction}


For transiting planets, the Rossiter-McLaughlin effect \citep{Holt:1893, Rossiter:1924p869, McLaughlin:1924p872, Queloz:2000p247, Gaudi:2007p1507},  allows the measure of $\beta$ (also called $\lambda$ in the literature), which is the projection on the sky of the obliquity $\psi$, between the stellar spin axis and the orbital spin axis.

Up until recently planets were thought to be mostly on orbits coplanar with their star's equator \citep{Fabrycky:2009p1845}, something in line with predictions of disc migration \citep{Lin:1996p5847, Ward:1997p11274}. More recently a number of  papers have shown that hot Jupiters on non coplanar orbits are common, including some planets on retrograde orbits \citep{Hebrard:2008p226, Moutou:2009p2007, Narita:2009p5188, Winn:2009p3712, Anderson:2010p5177, Queloz:2010p7376, Triaud:2010p8039}. Those measurements have been interpreted as showing that dynamical events are probably not uncommon and that not all systems can be understood by disc migration alone. Strong dynamical events such as planet-planet scattering \citep{Rasio:1996p3680, Juric:2008p2882, Chatterjee:2008p2971}, or more secular processes such as Kozai-Lidov oscillations \citep{Wu:2007p4179, Fabrycky:2007p3141, Nagasawa:2008p2997, Naoz:2011p8938}, or chaotic interactions \citep{Wu:2011p14507} would place a planet on a highly eccentric orbit, whose passage at periastron is sufficiently close that tidal dissipation causes the planet to lose angular momentum and circularise around its star.\\

Understanding the origin of hot Jupiters is one of the keys to shedding light onto the processes that act during planet formation as well as those acting after planets have formed. Those processes allow us to place constraints on what happened and did not happen in our own Solar System. They will also help us match more accurately theoretical predictions of planet formation done in population synthesis simulations to the parameter space that planets currently occupy, as given by the observations (eg. \citet{Ida:2004p11614} and \citet{Mordasini:2009p8294}). \\

\citet{Matsumura:2010p8927} remark that if misaligned hot Jupiters do not require disc migration, aligned planets are not in contradiction with a scenario involving dynamical interactions and tidal migration, as planets will tend to realign with the star (see also \citet{Hut:1981p2945}  and \citet{Barker:2009p11693}).


\citet{Winn:2010p7311} point out a correlation between the stellar effective temperature and the spin/orbit angle. For stars with $T_\mathrm{eff} > 6250$\,K, fewer aligned systems are found compared to stars with lower effective temperatures. This would show that tidal realignment timescales are different for different stars, as proposed by \citet{Zahn:1977p14439} in the context of binaries. 
\citet{Schlaufman:2010p7946} presents an independent confirmation of that correlation, using a different methodology. \\



The aim of this letter is to combine the observational facts and offer an explanation. The results will then be discuss  in light of the currently available theoretical framework.

\section{Motivation}
\begin{figure}
\centering
\includegraphics[width=8.cm]{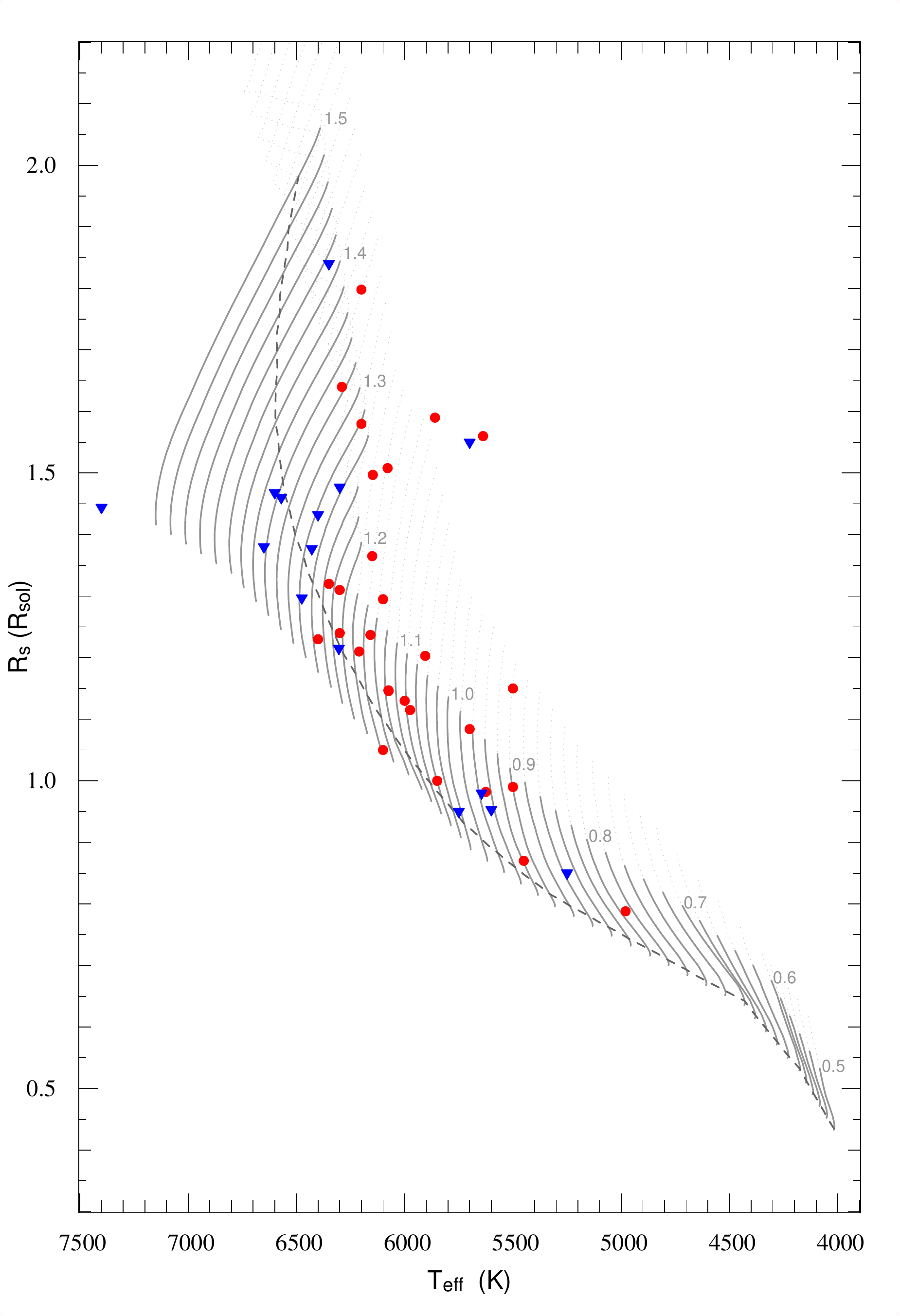}
\caption{Main Sequence showing the Geneva stellar evolution tracks for solar metallicity as presented in Mowlavi et al. (submitted) and plotted using $R_\star$ (in $R_\odot$) as a function of $T_\mathrm{eff}$. Tracks are labelled in units of $M_\odot$. Dashed line show the 2 Gyr isochrone. Overplotted are the systems for which we have  Rossiter-McLaughlin measurements. Aligned systems are red circles, misaligned systems are blue triangles. Higher metallicities will move the tracks to the right. Data obtained from {\sc Exoplanet.eu}}
\label{fig:tracks}
\end{figure}

The lack of aligned systems for stars with $T_\mathrm{eff} > 6250$\,K that is noticed in \citet{Winn:2010p7311} could also be explained by stellar physics combined with an observational bias:
as predicted by stellar evolution, stars with masses greater than about $\,1.2\,M_\odot$ start on the Zero Age Main Sequence with temperatures higher than $6250$\,K. When H-core burning stops, they have cooled by several hundred Kelvin (fig. \ref{fig:tracks}). They do so in 3 to 4 Gyrs. This means that, while the planet and the star progressively realign, the star itself cools down. We are thus left with an aligned planet around an older, cooler star. Some, more massive, stars will cool to temperatures above 6250\,K, but the timescale for realignment might be longer than the Main Sequence lifetime. Once they leave the Main Sequence, stars becomes too large for planets to be discovered by ground-based transit surveys as the contrast becomes too small.
We thus have a bias to see misaligned planets around hot stars, notably because we may not detect their aligned population. 

This explanation could be combined to the different realignment timescales described in \citet{Winn:2010p7311} and \citet{Zahn:1977p14439} since, as the star ages and cools, its convective zone would become larger too. 
If that explanation is right, we should expect a correlation between stellar age and alignment. 

The average stellar density, $\rho_\star$, is obtained directly from the planetary transit signal \citep{Sozzetti:2007p2647}, the effective temperature, $T_\mathrm{eff}$, and metallicity, $Z$, can be obtained via spectral analysis. Stellar mass and stellar age can be estimated from interpolating the stellar evolution tracks in $(\rho_\star, T_\mathrm{eff}, Z)$ space. Interestingly, stars $>1.2\,M_\odot$ spend less time on the Main Sequence, but increase their radii more than solar mass stars do. We thus have a higher resolution on the tracks to estimate ages on more massive stars than on solar mass stars. Such a subsample should give the most precise and accurate ages that we can get. This is the sample used in this letter.

\section{Sample selection}

Let us take only the most secure measurements for the projected spin/orbit angle\footnote{Some measurements have been omitted for the following reasons: CoRoT-3 (sampling is poor \citep{Triaud:2009p7520}), CoRoT-11 and Kepler-8 (transits are incomplete \citep{Gandolfi:2010p11112,Jenkins:2010p11060}) and WASP-1 (angle is unsure \citep{Albrecht:2011p14546}).}, for planets with stars $\geq 1.2\,M_\odot$. There are 22 objects in the sample (table \ref{tab:Param}). The sample is divided in two: stars $\geq 1.3\,M_\odot$ (8 stars) and stars between $1.2$ and $1.3\,M_\odot$ (14 stars). The angle and age estimates were obtained from the literature, but for WASP-17, whose error bar on the age was large. It was re-estimated for this letter, using the stellar parameters and density presented in \citet{Triaud:2010p8039} and interpolating in the Geneva tracks (Mowlavi et al. submitted). The new age estimate is $2.3\pm0.6$ Gyr. Its error bar is consistent with age measurements made by other teams. The new value is presented along with all other values in table \ref{tab:Param}.

\begin{figure}
\centering
\includegraphics[width=8.5cm]{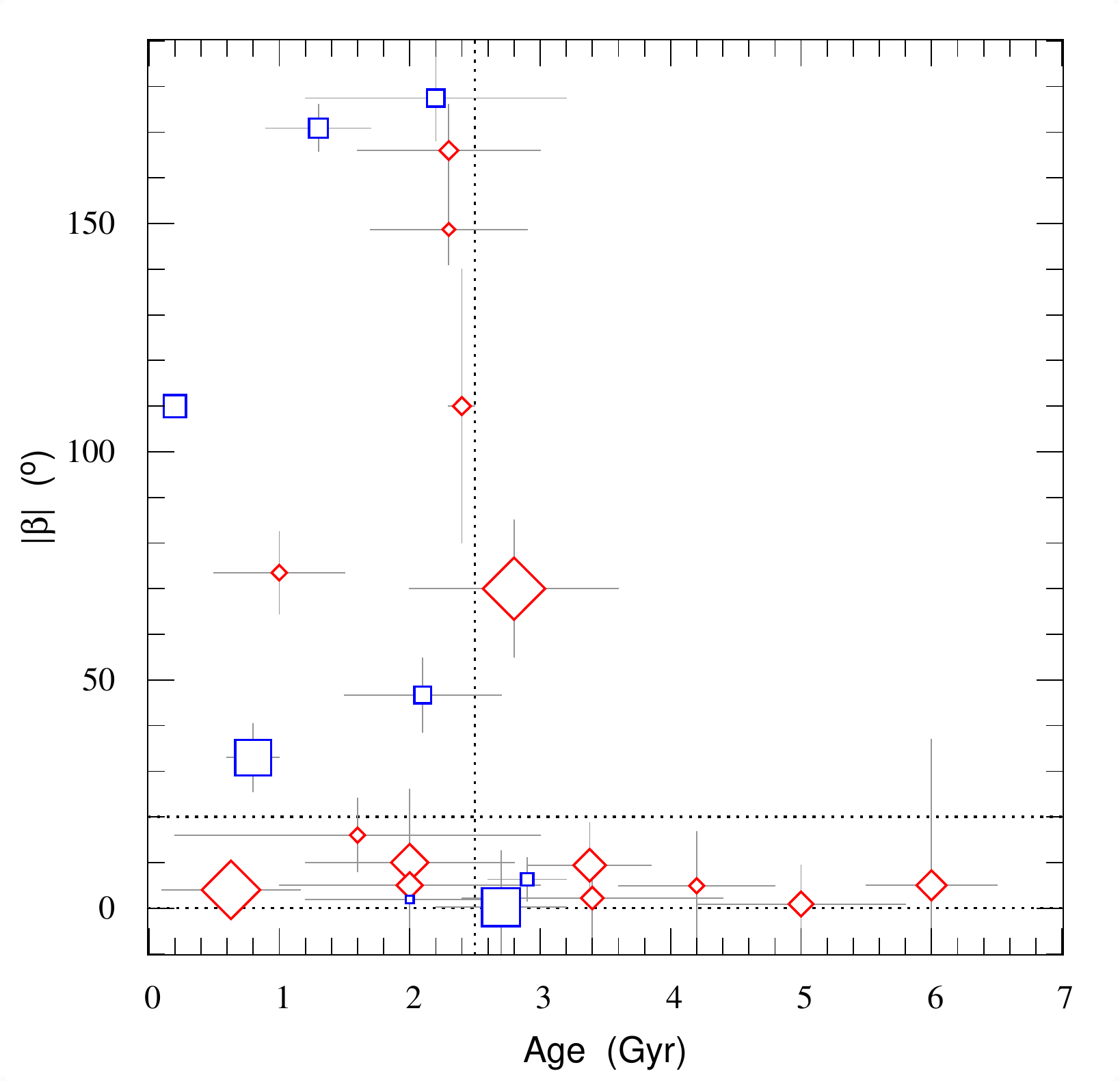}
\caption{Secure, absolute values of $\beta$ against stellar age (in Gyr), for stars with $M_\star \geq 1.2\,M_\odot$. Size of the symbols scales with planet mass. In blue squares, stars with $M_\star \geq 1.3\,M_\odot$; in red diamonds $1.3 > M_\star \geq 1.2\,M_\odot$. Horizontal dotted line show where aligned systems are. Vertical dotted line shows the age at which where misaligned planets start to disappear.}
\label{fig:betaAge}
\end{figure}

\begin{figure}
\centering
\includegraphics[width=8.5cm]{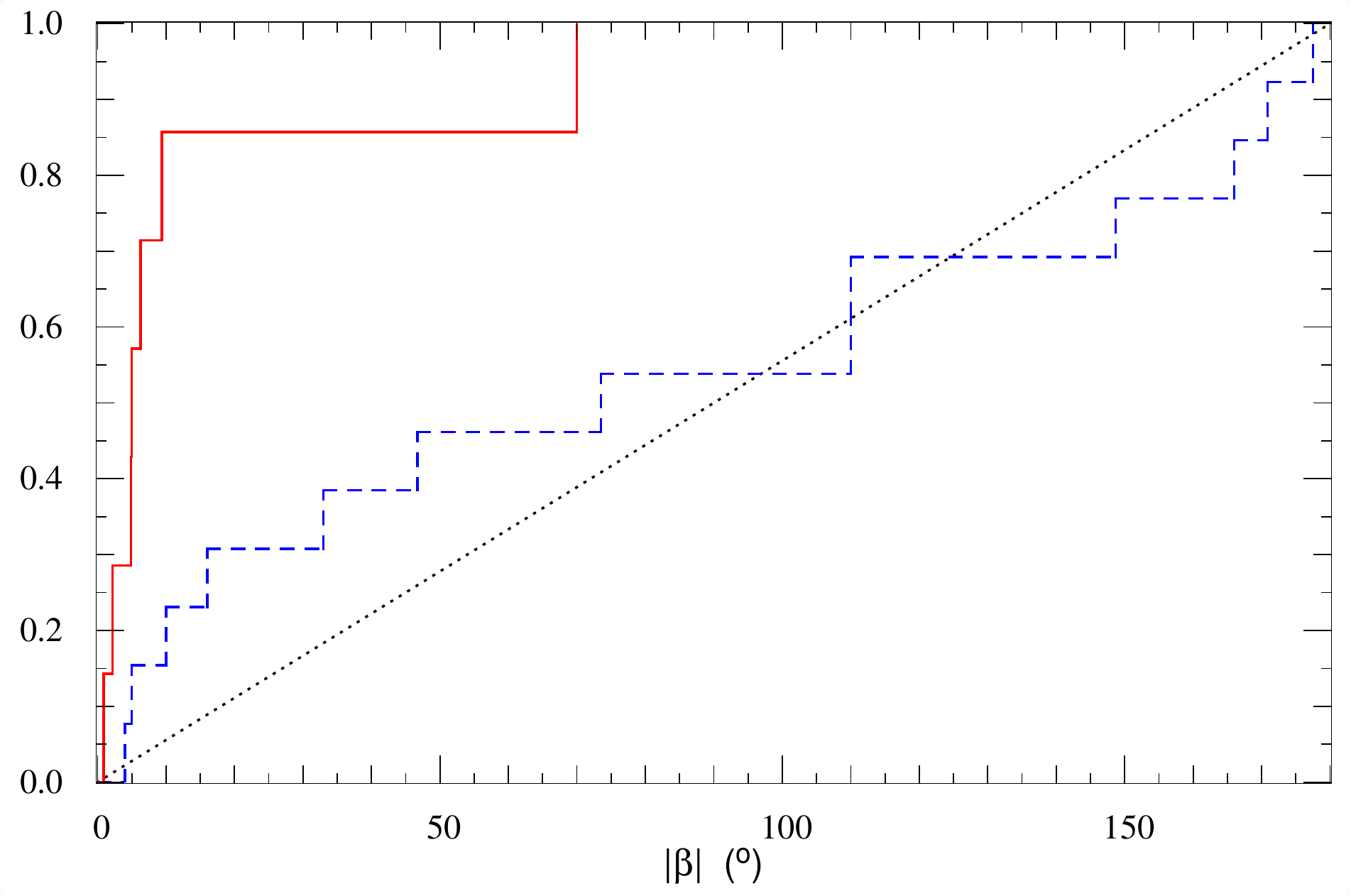}
\caption{Cumulative distributions in orbital inclinations for systems younger than 2.5 Gyr (dashed blue), and older (plain red). For comparison, a uniform distribution (dotted black).}
\label{fig:betaAgeCumul}
\end{figure}

Plotting the absolute values of the measured projected spin/orbit angle $\beta$ against stellar age (fig. \ref{fig:betaAge}), a pattern is obvious and as sharp as that presented in \citet{Winn:2010p7311}. While observationally, there should be no bias to preferentially detect aligned systems instead of misaligned systems at any age, stars older than $\sim$ 2.5\,Gyr show mostly aligned systems (rms = $22{^\circ}$, median = $5{^\circ}$). For stars that are younger we have a large range of obliquities (rms = $66^\circ$, median = $60{^\circ}$).  Figure \ref{fig:betaAgeCumul} displays the cumulative distributions on either side of the 2.5 Gyr age limit.

To test the robustness of the pattern, a Monte Carlo simulation was performed taking the data with ages $< 2.5$ Gyr as a fiducial zone from which random samples of 8 measurements were drawn, allowing for repetitions. There is  $<$ 4\% chance to draw a sample with median $< 10{^\circ}$ and rms $< 60{^\circ}$ which would allow a sample having seven aligned systems and one retrograde system. If restricting the rms within $30{^\circ}$, similar to that observed, there is a probability $<$ 1\% that the distributions on either side of the 2.5 Gyr age are the same. Drawing randomly from the overall sample, there is a 2.6\% chance to obtain a cluster containing 7 aligned systems and another at any angle $> 20^\circ$. In addition a Kolmogorov-Smirnov test was carried out, also comparing the distribution in $\beta$ on either side of the 2.5 Gyr limit. A $D = 0.661$ is obtained corresponding to a probability of 1.2\% that both distributions are the same\footnote{the same test on the pattern presented in \citet{Winn:2010p7311} gives 6.1\% chance that the distributions on either side of 6250\,K are the same.}. The same test shows that the distribution of angles around stars younger than 2.5 Gyr has about 22\% chance to be compatible with a uniform distribution, while for the older sample, this chance is of order $10^{-5}$. By rearranging the data, selecting various cut-off and computing the KS test at each step, the probability of having two such different populations arising by chance is estimated to about 7\%.  It can be affirmed there is tentative evidence of a pattern in the data.\\ 
We see that stars with masses $\geq 1.3\,M_\odot$ are all younger than 3\,Gyr. Thus, when observing few aligned systems on stars with $T_\mathrm{eff} > 6250\,K$, \citet{Winn:2010p7311} were in fact detecting an effect  due to stellar age, or rather, time since planet formation. \\

Like for all multivariate problems, figure \ref{fig:betaAge} offers an incomplete picture: it only shows two quantities in relation with time. At the moment orbital separations and mass ratios are quite similar since the bulk of the discoveries have been done by ground-based transit search programs. With increasing numbers of measurements over a larger parameter space we will eventually need to account for those extra parameters.

\section{Discussion}


The large variety of angles around the younger stars suggests that some misaligning mechanism happens during the youth of planetary systems. Notably, in combination with results by \citet{Watson:2011p11844} showing no evidence for misaligned protoplanetary discs, it lends strong support to a planet-planet scattering scenario occurring during the last stages of planet formation or soon in the aftermath of the disc dispersal like described in \citet{Matsumura:2010p8930}.

When preparing figure \ref{fig:betaAge}, reason dictated that a dearth of old, misaligned systems was expected, not an absence. The complete lack of misaligned planets orbiting stars older than 2.5\,Gyr in the current sample came somewhat as a surprise as secular interactions could place planets on inclined orbits well after the disc dissipated. A system presenting such characteristics can be found among the "older" systems: HAT-P-13, whose current configuration may have originated from secular interactions \citep{Mardling:2010p12652}. If that history is right, its observed coplanarity may be a chance alignment. Chance alignments can occur easily since firstly, we observe a projected angle, $\beta$, and not the real obliquity $\psi$ and secondly, theoretical predictions such as \citet{Wu:2007p4179}, \citet{Fabrycky:2007p3141} and \citet{Nagasawa:2008p2997} predict very high orbital inclinations, but also a number of aligned systems. 


There is great interest in matching those theoretical distributions to observations (notably for young hot Jupiters), but the evolving nature of the spin/orbit angle distribution makes this a tricky task. Multi-body dynamics are less concerned about absolute masses than about mass ratio. In systems where no Jupiter has formed, we would expect planet-planet scattering between Neptune-mass planets producing an inclined hot Neptune population. If the inital stages will be similar, the later ones will not: tidal circularisation and realignment timescales will be different. Spin/orbit angles for planets of masses $< 0.1 \,M_\mathrm{Jup}$ will be less affected by tidal realignment and offer a closer picture of the initial spin/orbit angle distribution than hot Jupiters. A hot Neptune, Hat-P-11\,b has been recently detected misaligned by \citet{Winn:2010p8446} and confirmed by \citet{SanchisOjeda:2011p12868}.

This work has focused on stars with masses $\geq 1.2\,M_\odot$. 
If age is what determines primarily whether a hot Jupiter is observed aligned or misaligned, since solar mass stars are detected in average older than more massive stars, it is not surprising that their planets are coplanar. There nevertheless is an interest in looking at that population carefully which stems from work by \citet{Burkert:2007p12867}, \citet{Currie:2009p12859} and \citet{Alibert:2011p11846} who argue that discs around the more massive stars are not long lived enough to produce an aligned hot Jupiter population via disc migration. In the mean time, if planet formation is more efficient in more massive discs (found around more massive stars), then one could expect a higher occurrence of planet-planet scattering around such stars. If this is true, it could point towards two pathways for bringing hot Jupiters to their observed location which would be dependent on stellar mass. 
Unfortunately stellar ages are less precisely determined for solar mass stars as illustrated by the isochrone on figure \ref{fig:tracks}.\\

The change in the shape of the distribution of spin/orbit angles with time is indicative of some orbital evolution, presumably through tidal interactions between the star and the planet. \citet{Barker:2009p11693} show that retrograde planets decay into their star on timescales two to three times shorter than prograde planets would do, for given initial conditions. Their infall timescale for a typical, retrograde, hot Jupiter are of order of a few Gyrs. \citet{Winn:2010p7311} present similar behaviour. In addition they show that, for a given stellar mass, a more massive planet will realign and in-spiral faster than a lighter one\footnote{Incidentally, this could explain a second observed feature in relation to the angle $\beta$. As shown in \citet{Moutou:2011p14504}, there is a lack of retrograde massive planets ($> 5\,M_\mathrm{Jup}$), something expected if retrograde, massive planets realign faster to their star than other planets do.}. 
In both papers the retrograde planets realign with the star but only shortly before falling into it. It would thus be unlikely to observe them at these very particular phases. Nevertheless such examples could be found in WASP-12, 18 and 19 (eg. \citet{Hellier:2011p10960}). \citet{Matsumura:2010p8927} describe how planets initially placed on mildly inclined or aligned orbits, are less likely to in-fall  and  more likely to survive until observed. Nevertheless, in most cases tidal realignment corresponds to the disappearance of the planet. 

If retrograde planets plunge into their star as they tidally realign, a decreasing number of hot Jupiters should be observed with time.  
No such decreasing trend can be found when considering all the hot Jupiters presented in the literature around stars in the mass range considered for this paper. This is at odds with evidence of a trend between semimajor axis and stellar age showing a lack of very short orbits around older stars as presented by \citet{Jackson:2009p13324}  who interpreted it as evidence of the destructive tidal orbital decay of hot Jupiters. Looking at the semimajor axes of the targets in table \ref{tab:Param}, a similar trend appears. This may indicate we do not have enough objects yet to detect the expected decreasing fraction of hot Jupiters with time. Alternatively the results presented here, and those of \citet{Winn:2010p7311}, could be seen as evidence that realignment occurs faster than orbital decay. In a very recent development, \citet{Lai:2011p14503} shows how planets could realign their orbit faster than their semimajor axis would decay. \\

Finding out about the ultimate fate of hot Jupiters is of great interest and a subject of intense on-going research, fraught with challenges. For example, the tidal circularisation and realignment timescales notably depend on the orbital obliquity $\psi$, the ratio of masses, the scaled radius ($a/R_\star$) and the tidal quality factors, in the planet $Q'_\mathrm{p}$ and in the star, $Q'_\star$ \citep{Hut:1981p2945,Barker:2009p11693}. Most of the theoretical work currently assumes constant $Q'$  values when \citet{Ogilvie:2004p11737} showed they depend on the tidal frequency. \citet{Lai:2011p14503} indicates $Q'$ could vary for different physical processes. Similarly $R_\star$ is often assumed constant when clearly, in figure \ref{fig:tracks} a $1.3\,M_\odot$ star increases its radius by about 30\% in about 4 Gyrs.  \\


Stellar age estimates are notoriously difficult to obtain. The estimates that have been used here have been extracted by a variety of authors using different techniques on different sets of evolution models. The pattern resisted a blurring caused by systematic effects, displaying a certain robustness. 
Nevertheless, this letter should also be an incentive to continue obtaining Rossiter-McLaughlin measurements as well as check those stellar ages and derive them in a uniform manner. Similarly, accurate and precise age estimates for solar mass stars are dearly needed. One can access those via good determination of stellar parameters, using higher resolution spectroscopy for the $T_\mathrm{eff}$ and $Z$, and high precision photometry which will give $\rho_\star$. 
Stellar ages can also be estimated from asteroseismologic timeseries underlying the interest in having a planet-finding space mission with such capacity, like the proposed PLATO. Astrometric distance measurements from the GAIA satellite will soon give us an independent access to stellar radii.

\begin{table*}
\caption{Stellar and planetary parameters used to create figure \ref{fig:betaAge}. Values obtained using {\sc Exoplanet.eu}. The values of $\beta$ here are the absolute values. Error bars for the age are consistently the lower error bars presented in papers.}\label{tab:Param}
\begin{tabular}{lcccccl}
\hline
\hline
Name & $M_\star$ ($M_\odot$) &  $R_\star$  ($R_\odot$) &  $\beta$     ($^\circ$)     & Age  (Gyr) & $M_\mathrm{p} $ ($M_\mathrm{Jup}$) &    references\\
\hline
HAT-P-2& 1.36 & 1.64 & $0.2\pm12.3$& $2.7\pm0.5$ & 8.7   & \citep{Pal:2010p12466, Loeillet:2008p881}\\
HAT-P-4& 1.26 & 1.59 & $4.9\pm11.9$&$4.2\pm0.6$ & 0.7   & \citep{Kovacs:2007p12473, Winn:2011p11042}\\
HAT-P-6& 1.29 & 1.46 & $166\pm10$ & $2.3\pm0.7$  & 1.1   & \citep{Noyes:2008p12477,Hebrard:2011p11123}\\
HAT-P-7& 1.47 & 1.84 & $178\pm9$   &$2.2\pm1.0$& 1.8  & \citep{Pal:2008p4127,Winn:2009p3712}\\
HAT-P-8& 1.28 & 1.58 & $2.2\pm10.5$ & $3.4\pm1.0$ & 1.5  &\citep{Latham:2009p12478, Moutou:2011p14504}\\
HAT-P-9& 1.28	& 1.32 & $16\pm8$      & $1.6\pm1.4$   & 0.7  &\citep{Shporer:2009p12233,Moutou:2011p14504}\\
HAT-P-13&1.22&1.56 & $0.9\pm8.5$  & $5.0\pm0.8$   & 1.9  & \citep{Bakos:2009p12482,Winn:2010p11036}\\
HAT-P-14&1.38&1.47 & $171\pm5$    & $1.3\pm0.4$  & 2.2   & \citep{Torres:2010p12511,Winn:2011p11042}\\
HAT-P-16&1.22&1.24 & $10\pm16$   & $2.0\pm0.8$      & 4.2   & \citep{Buchhave:2010p12531, Moutou:2011p14504}\\
HAT-P-30&1.24&1.22 & $74\pm9$     & $1.0\pm0.5$      &0.7    & \citep{Johnson:2011p11065}    \\
HD\,17156&1.28&1.51&$9.4\pm9.3$ & $3.4\pm0.5$  &3.2     &\citep{Narita:2009p8532, Nutzman:2011p12545}\\
HD\,149026&1.30&1.50&$1.9\pm6.1$&$2.0\pm0.8$    & 0.4    &\citep{Sato:2005p2510,Wolf:2007p1867}\\
TrES\,4 & 1.39 & 1.80 & $6.3\pm4.7$ & $2.9\pm0.3$  & 0.9    & \citep{Narita:2010p7141, Chan:2011p12551} \\
WASP-3&1.23 & 1.31 & $5\pm5$        & $ 2.0\pm1.0$       & 2.1    & \citep{Pollacco:2008p5241,Miller:2010p8226}\\
WASP-7&1.28 & 1.43 & $110\pm30$ &$2.4\pm0.1$    & 1.0  & \citep[Triaud et al. in prep]{Southworth:2011p12650}\\
WASP-14&1.32&1.30& $33\pm7$      &$0.8\pm0.2$   &7.7     & \citep{Joshi:2009p2063, Johnson:2009p3754}\\
WASP-17&1.20&1.38& $149\pm8$    & $2.3\pm0.6$  &0.5     & \citep{Triaud:2010p8039}; this paper\\
WASP-18&1.24&1.23& $4\pm5$         &$0.6\pm0.5$ & 10.4    & \citep{Hellier:2009p3885,Triaud:2010p8039}\\
WASP-33&1.50&1.44& $110\pm0.3$ &$0.2\pm0.2$ & $<4$       & \citep{CollierCameron:2010p11163}\\
WASP-38&1.22&1.37& $ 5\pm32 $     &$6.0\pm0.5 $   & 2.7      &\citep{Barros:2011p8210,Simpson:2011p8894}\\
XO-3     & 1.21 & 1.38 & $70\pm15    $&$2.8\pm0.8  $ & 11.8    &\citep{JohnsKrull:2008p2576,Winn:2009p3777}\\
XO-4     & 1.32 & 1.55 & $47\pm8    $     &$2.1\pm0.6$    &1.7      &\citep{McCullough:2008p12585,Narita:2010p8435}\\
\hline
\end{tabular}
\end{table*}

\begin{acknowledgements} 
Many thanks are given to those that gave feedback on this work and those with whom the many discussions helped me in producing this work, especially Rosemary Mardling, Soko Matsumura and Johannes Sahlmann, but also Christoph Mordasini, Pedro Figueira, Josh Winn, Andrew Collier Cameron, Georges Meynet, Micha\"el Gillon  and Damien S\'egransan. Thanks to Nami Mowlavi and the Geneva stellar evolution group for preparing stellar tracks to my specifications. Eternal recognition also go to Didier Queloz, my PhD supervisor for teaching me how to research while giving me freedom and independence. I wish to convey thanks to my editor, Tristan Guillot, and to an anonymous referee for refining this work and understanding it. I would also like to acknowledge the use of Jean Schneider's {\sc exoplanet.eu} and Ren\'e Heller's encyclopaedia for the Rossiter-McLaughlin effect. This work is supported by the Swiss Fond National de Recherche Scientifique. 
\end{acknowledgements} 

\bibliographystyle{aa}
\bibliography{bibtex}

\begin{thebibliography}{82}
\expandafter\ifx\csname natexlab\endcsname\relax\def\natexlab#1{#1}\fi

\bibitem[{Albrecht {et~al.}(2011)Albrecht, Winn, Johnson, Butler, Crane,
  Shectman, Thompson, Narita, Sato, Hirano, Enya, \&
  Fischer}]{Albrecht:2011p14546}
Albrecht, S., Winn, J.~N., Johnson, J.~A., {et~al.} 2011, ApJ, 738, 50

\bibitem[{Alibert {et~al.}(2011)Alibert, Mordasini, \&
  Benz}]{Alibert:2011p11846}
Alibert, Y., Mordasini, C., \& Benz, W. 2011, A{\&}A, 526, 63

\bibitem[{Anderson {et~al.}(2010)Anderson, Hellier, Gillon, Triaud, Smalley,
  Hebb, Cameron, Maxted, Queloz, West, Bentley, Enoch, Horne, Lister, Mayor,
  Parley, Pepe, Pollacco, S{\'e}gransan, Udry, \& Wilson}]{Anderson:2010p5177}
Anderson, D.~R., Hellier, C., Gillon, M., {et~al.} 2010, ApJ, 709, 159

\bibitem[{Bakos {et~al.}(2009)Bakos, Howard, Noyes, Hartman, Torres,
  Kov{\'a}cs, Fischer, Latham, Johnson, Marcy, Sasselov, Stefanik, Sip{\H o}cz,
  Kov{\'a}cs, Esquerdo, P{\'a}l, L{\'a}z{\'a}r, Papp, \&
  S{\'a}ri}]{Bakos:2009p12482}
Bakos, G.~{\'A}., Howard, A.~W., Noyes, R.~W., {et~al.} 2009, ApJ, 707, 446

\bibitem[{Barker \& Ogilvie(2009)}]{Barker:2009p11693}
Barker, A.~J. \& Ogilvie, G.~I. 2009, MNRAS, 395, 2268

\bibitem[{Barros {et~al.}(2011)Barros, Faedi, Cameron, Lister, McCormac,
  Pollacco, Simpson, Smalley, Street, Todd, Triaud, Boisse, Bouchy,
  H{\'e}brard, Moutou, Pepe, Queloz, Santerne, Segransan, Udry, Bento, Butters,
  Enoch, Haswell, Hellier, Keenan, Miller, Moulds, Norton, Parley, Skillen,
  Watson, West, \& Wheatley}]{Barros:2011p8210}
Barros, S. C.~C., Faedi, F., Cameron, A.~C., {et~al.} 2011, A{\&}A, 525, 54

\bibitem[{Buchhave {et~al.}(2010)Buchhave, Bakos, Hartman, Torres, Kov{\'a}cs,
  Latham, Noyes, Esquerdo, Everett, Howard, Marcy, Fischer, Johnson, Andersen,
  F{\H u}r{\'e}sz, Perumpilly, Sasselov, Stefanik, B{\'e}ky, L{\'a}z{\'a}r,
  Papp, \& S{\'a}ri}]{Buchhave:2010p12531}
Buchhave, L.~A., Bakos, G.~{\'A}., Hartman, J.~D., {et~al.} 2010, ApJ, 720,
  1118

\bibitem[{Burkert \& Ida(2007)}]{Burkert:2007p12867}
Burkert, A. \& Ida, S. 2007, ApJ, 660, 845

\bibitem[{Cameron {et~al.}(2010)Cameron, Guenther, Smalley, McDonald, Hebb,
  Andersen, Augusteijn, Barros, Brown, Cochran, Endl, Fossey, Hartmann, Maxted,
  Pollacco, Skillen, Telting, Waldmann, \& West}]{CollierCameron:2010p11163}
Cameron, A.~C., Guenther, E., Smalley, B., {et~al.} 2010, MNRAS, 407, 507

\bibitem[{Chan {et~al.}(2011)Chan, Ingemyr, Winn, Holman, Sanchis-Ojeda,
  Esquerdo, \& Everett}]{Chan:2011p12551}
Chan, T., Ingemyr, M., Winn, J.~N., {et~al.} 2011, AJ, 141, 179

\bibitem[{Chatterjee {et~al.}(2008)Chatterjee, Ford, Matsumura, \&
  Rasio}]{Chatterjee:2008p2971}
Chatterjee, S., Ford, E.~B., Matsumura, S., \& Rasio, F.~A. 2008, ApJ, 686, 580

\bibitem[{Currie(2009)}]{Currie:2009p12859}
Currie, T. 2009, ApJ, 694, L171

\bibitem[{Fabrycky \& Tremaine(2007)}]{Fabrycky:2007p3141}
Fabrycky, D. \& Tremaine, S. 2007, ApJ, 669, 1298

\bibitem[{Fabrycky \& Winn(2009)}]{Fabrycky:2009p1845}
Fabrycky, D.~C. \& Winn, J.~N. 2009, ApJ, 696, 1230

\bibitem[{Gandolfi {et~al.}(2010)Gandolfi, H{\'e}brard, Alonso, Deleuil,
  Guenther, Fridlund, Endl, Eigm{\"u}ller, Csizmadia, Havel, Aigrain, Auvergne,
  Baglin, Barge, Bonomo, Bord{\'e}, Bouchy, Bruntt, Cabrera, Carpano, Carone,
  Cochran, Deeg, Dvorak, Eisl{\"o}ffel, Erikson, Ferraz-Mello, Gazzano, Gibson,
  Gillon, Gondoin, Guillot, Hartmann, Hatzes, Jorda, Kabath, L{\'e}ger,
  Llebaria, Lammer, MacQueen, Mayor, Mazeh, Moutou, Ollivier, P{\"a}tzold,
  Pepe, Queloz, Rauer, Rouan, Samuel, Schneider, Stecklum, Tingley, Udry, \&
  Wuchterl}]{Gandolfi:2010p11112}
Gandolfi, D., H{\'e}brard, G., Alonso, R., {et~al.} 2010, A{\&}A, 524, 55

\bibitem[{Gaudi \& Winn(2007)}]{Gaudi:2007p1507}
Gaudi, B.~S. \& Winn, J.~N. 2007, ApJ, 655, 550

\bibitem[{H{\'e}brard {et~al.}(2008)H{\'e}brard, Bouchy, Pont, Loeillet, Rabus,
  Bonfils, Moutou, Boisse, Delfosse, Desort, Eggenberger, Ehrenreich,
  Forveille, Lagrange, Lovis, Mayor, Pepe, Perrier, Queloz, Santos,
  S{\'e}gransan, Udry, \& Vidal-Madjar}]{Hebrard:2008p226}
H{\'e}brard, G., Bouchy, F., Pont, F., {et~al.} 2008, A{\&}A, 488, 763

\bibitem[{H{\'e}brard {et~al.}(2011)H{\'e}brard, Ehrenreich, Bouchy, Delfosse,
  Moutou, Arnold, Boisse, Bonfils, D{\'\i}az, Eggenberger, Forveille, Lagrange,
  Lovis, Pepe, Perrier, Queloz, Santerne, Santos, S{\'e}gransan, Udry, \&
  Vidal-Madjar}]{Hebrard:2011p11123}
H{\'e}brard, G., Ehrenreich, D., Bouchy, F., {et~al.} 2011, A{\&}A, 527, L11

\bibitem[{Hellier {et~al.}(2009)Hellier, Anderson, Cameron, Gillon, Hebb,
  Maxted, Queloz, Smalley, Triaud, West, Wilson, Bentley, Enoch, Horne, Irwin,
  Lister, Mayor, Parley, Pepe, Pollacco, Segransan, Udry, \&
  Wheatley}]{Hellier:2009p3885}
Hellier, C., Anderson, D.~R., Cameron, A.~C., {et~al.} 2009, Nature, 460, 1098

\bibitem[{Hellier {et~al.}(2011)Hellier, Anderson, Collier-Cameron, Miller,
  Queloz, Smalley, Southworth, \& Triaud}]{Hellier:2011p10960}
Hellier, C., Anderson, D.~R., Collier-Cameron, A., {et~al.} 2011, ApJ, 730, L31

\bibitem[{Holt(1893)}]{Holt:1893}
Holt, J. 1893, Astronomy {\&} Astrophysics, XII, 646

\bibitem[{Hut(1981)}]{Hut:1981p2945}
Hut, P. 1981, A{\&}A, 99, 126

\bibitem[{Ida \& Lin(2004)}]{Ida:2004p11614}
Ida, S. \& Lin, D. N.~C. 2004, ApJ, 604, 388

\bibitem[{Jackson {et~al.}(2009)Jackson, Barnes, \&
  Greenberg}]{Jackson:2009p13324}
Jackson, B., Barnes, R., \& Greenberg, R. 2009, ApJ, 698, 1357

\bibitem[{Jenkins {et~al.}(2010)Jenkins, Borucki, Koch, Marcy, Cochran, Welsh,
  Basri, Batalha, Buchhave, Brown, Caldwell, Dunham, Endl, Fischer, Gautier,
  Geary, Gilliland, Howell, Isaacson, Johnson, Latham, Lissauer, Monet, Rowe,
  Sasselov, Howard, MacQueen, Orosz, Chandrasekaran, Twicken, Bryson, Quintana,
  Clarke, Li, Allen, Tenenbaum, Wu, Meibom, Klaus, Middour, Cote, McCauliff,
  Girouard, Gunter, Wohler, Hall, Ibrahim, Uddin, Wu, Bhavsar, Cleve, Pletcher,
  Dotson, \& Haas}]{Jenkins:2010p11060}
Jenkins, J.~M., Borucki, W.~J., Koch, D.~G., {et~al.} 2010, ApJ, 724, 1108

\bibitem[{Johns-Krull {et~al.}(2008)Johns-Krull, McCullough, Burke, Valenti,
  Janes, Heasley, Prato, Bissinger, Fleenor, Foote, Garcia-Melendo, Gary,
  Howell, Mallia, Masi, \& Vanmunster}]{JohnsKrull:2008p2576}
Johns-Krull, C.~M., McCullough, P.~R., Burke, C.~J., {et~al.} 2008, ApJ, 677,
  657

\bibitem[{Johnson {et~al.}(2009)Johnson, Winn, Albrecht, Howard, Marcy, \&
  Gazak}]{Johnson:2009p3754}
Johnson, J.~A., Winn, J.~N., Albrecht, S., {et~al.} 2009, PASP, 121, 1104

\bibitem[{Johnson {et~al.}(2011)Johnson, Winn, Bakos, Hartman, Morton, Torres,
  Kov{\'a}cs, Latham, Noyes, Sato, Esquerdo, Fischer, Marcy, Howard, Buchhave,
  F{\H u}r{\'e}sz, Quinn, B{\'e}ky, Sasselov, Stefanik, L{\'a}z{\'a}r, Papp, \&
  S{\'a}ri}]{Johnson:2011p11065}
Johnson, J.~A., Winn, J.~N., Bakos, G.~{\'A}., {et~al.} 2011, ApJ, 735, 24

\bibitem[{Joshi {et~al.}(2009)Joshi, Pollacco, Cameron, Skillen, Simpson,
  Steele, Street, Stempels, Christian, Hebb, Bouchy, Gibson, H{\'e}brard,
  Keenan, Loeillet, Meaburn, Moutou, Smalley, Todd, West, Anderson, Bentley,
  Enoch, Haswell, Hellier, Horne, Irwin, Lister, McDonald, Maxted, Mayor,
  Norton, Parley, Perrier, Pont, Queloz, Ryans, Smith, Udry, Wheatley, \&
  Wilson}]{Joshi:2009p2063}
Joshi, Y.~C., Pollacco, D., Cameron, A.~C., {et~al.} 2009, MNRAS, 392, 1532

\bibitem[{Juri{\'c} \& Tremaine(2008)}]{Juric:2008p2882}
Juri{\'c}, M. \& Tremaine, S. 2008, ApJ, 686, 603

\bibitem[{Kov{\'a}cs {et~al.}(2007)Kov{\'a}cs, Bakos, Torres, Sozzetti, Latham,
  Noyes, Butler, Marcy, Fischer, Fern{\'a}ndez, Esquerdo, Sasselov, Stefanik,
  P{\'a}l, L{\'a}z{\'a}r, Papp, \& S{\'a}ri}]{Kovacs:2007p12473}
Kov{\'a}cs, G., Bakos, G.~{\'A}., Torres, G., {et~al.} 2007, ApJ, 670, L41

\bibitem[{Lai(2011)}]{Lai:2011p14503}
Lai, D. 2011, arXiv, astro-ph.EP, 10 pages

\bibitem[{Latham {et~al.}(2009)Latham, Bakos, Torres, Stefanik, Noyes,
  Kov{\'a}cs, P{\'a}l, Marcy, Fischer, Butler, Sip{\H o}cz, Sasselov, Esquerdo,
  Vogt, Hartman, Kov{\'a}cs, L{\'a}z{\'a}r, Papp, \&
  S{\'a}ri}]{Latham:2009p12478}
Latham, D.~W., Bakos, G.~{\'A}., Torres, G., {et~al.} 2009, ApJ, 704, 1107

\bibitem[{Lin {et~al.}(1996)Lin, Bodenheimer, \& Richardson}]{Lin:1996p5847}
Lin, D. N.~C., Bodenheimer, P., \& Richardson, D.~C. 1996, Nature, 380, 606

\bibitem[{Loeillet {et~al.}(2008)Loeillet, Shporer, Bouchy, Pont, Mazeh,
  Beuzit, Boisse, Bonfils, da~Silva, Delfosse, Desort, Ecuvillon, Forveille,
  Galland, Gallenne, H{\'e}brard, Lagrange, Lovis, Mayor, Moutou, Pepe,
  Perrier, Queloz, S{\'e}gransan, Sivan, Santos, Tsodikovich, Udry, \&
  Vidal-Madjar}]{Loeillet:2008p881}
Loeillet, B., Shporer, A., Bouchy, F., {et~al.} 2008, A{\&}A, 481, 529

\bibitem[{Mardling(2010)}]{Mardling:2010p12652}
Mardling, R.~A. 2010, MNRAS, 407, 1048

\bibitem[{Matsumura {et~al.}(2010{\natexlab{a}})Matsumura, Peale, \&
  Rasio}]{Matsumura:2010p8927}
Matsumura, S., Peale, S.~J., \& Rasio, F.~A. 2010{\natexlab{a}}, ApJ, 725, 1995

\bibitem[{Matsumura {et~al.}(2010{\natexlab{b}})Matsumura, Thommes, Chatterjee,
  \& Rasio}]{Matsumura:2010p8930}
Matsumura, S., Thommes, E.~W., Chatterjee, S., \& Rasio, F.~A.
  2010{\natexlab{b}}, ApJ, 714, 194

\bibitem[{McCullough {et~al.}(2008)McCullough, Burke, Valenti, Long,
  Johns-Krull, Machalek, Janes, Taylor, Gregorio, Foote, Gary, Fleenor,
  Garc{\'\i}a-Melendo, \& Vanmunster}]{McCullough:2008p12585}
McCullough, P.~R., Burke, C.~J., Valenti, J.~A., {et~al.} 2008, eprint arXiv,
  0805, 2921, 27 pages, 7 figures, submitted to ApJ

\bibitem[{McLaughlin(1924)}]{McLaughlin:1924p872}
McLaughlin, D.~B. 1924, ApJ, 60, 22

\bibitem[{Miller {et~al.}(2010)Miller, Cameron, Simpson, Pollacco, Enoch,
  Gibson, Queloz, Triaud, H{\'e}brard, Boisse, Moutou, \&
  Skillen}]{Miller:2010p8226}
Miller, G. R.~M., Cameron, A.~C., Simpson, E.~K., {et~al.} 2010, A{\&}A, 523,
  52

\bibitem[{Mordasini {et~al.}(2009)Mordasini, Alibert, Benz, \&
  Naef}]{Mordasini:2009p8294}
Mordasini, C., Alibert, Y., Benz, W., \& Naef, D. 2009, A{\&}A, 501, 1161

\bibitem[{Moutou {et~al.}(2011)Moutou, D{\'\i}az, Udry, H{\'e}brard, Bouchy,
  Santerne, Ehrenreich, Arnold, Boisse, Bonfils, Delfosse, Eggenberger,
  Forveille, Lagrange, Lovis, Martinez, Pepe, Perrier, Queloz, Santos,
  S{\'e}gransan, Toublanc, Troncin, Vanhuysse, \&
  Vidal-Madjar}]{Moutou:2011p14504}
Moutou, C., D{\'\i}az, R.~F., Udry, S., {et~al.} 2011, A{\&}A, 533, 113

\bibitem[{Moutou {et~al.}(2009)Moutou, H{\'e}brard, Bouchy, Eggenberger,
  Boisse, Bonfils, Gravallon, Ehrenreich, Forveille, Delfosse, Desort,
  Lagrange, Lovis, Mayor, Pepe, Perrier, Pont, Queloz, Santos, S{\'e}gransan,
  Udry, \& Vidal-Madjar}]{Moutou:2009p2007}
Moutou, C., H{\'e}brard, G., Bouchy, F., {et~al.} 2009, A{\&}A, 498, L5

\bibitem[{Nagasawa {et~al.}(2008)Nagasawa, Ida, \& Bessho}]{Nagasawa:2008p2997}
Nagasawa, M., Ida, S., \& Bessho, T. 2008, ApJ, 678, 498

\bibitem[{Naoz {et~al.}(2011)Naoz, Farr, Lithwick, Rasio, \&
  Teyssandier}]{Naoz:2011p8938}
Naoz, S., Farr, W.~M., Lithwick, Y., Rasio, F.~A., \& Teyssandier, J. 2011,
  Nature, 473, 187

\bibitem[{Narita {et~al.}(2010{\natexlab{a}})Narita, Hirano, Sanchis-Ojeda,
  Winn, Holman, Sato, Aoki, \& Tamura}]{Narita:2010p8435}
Narita, N., Hirano, T., Sanchis-Ojeda, R., {et~al.} 2010{\natexlab{a}}, PASJ,
  62, L61

\bibitem[{Narita {et~al.}(2009{\natexlab{a}})Narita, Hirano, Sato, Winn, Suto,
  Turner, Aoki, Tamura, \& Yamada}]{Narita:2009p8532}
Narita, N., Hirano, T., Sato, B., {et~al.} 2009{\natexlab{a}}, PASJ, 61, 991

\bibitem[{Narita {et~al.}(2009{\natexlab{b}})Narita, Sato, Hirano, \&
  Tamura}]{Narita:2009p5188}
Narita, N., Sato, B., Hirano, T., \& Tamura, M. 2009{\natexlab{b}}, PASJ, 61,
  L35

\bibitem[{Narita {et~al.}(2010{\natexlab{b}})Narita, Sato, Hirano, Winn, Aoki,
  \& Tamura}]{Narita:2010p7141}
Narita, N., Sato, B., Hirano, T., {et~al.} 2010{\natexlab{b}}, PASJ, 62, 653

\bibitem[{Noyes {et~al.}(2008)Noyes, Bakos, Torres, P{\'a}l, Kov{\'a}cs,
  Latham, Fern{\'a}ndez, Fischer, Butler, Marcy, Sip{\H o}cz, Esquerdo,
  Kov{\'a}cs, Sasselov, Sato, Stefanik, Holman, L{\'a}z{\'a}r, Papp, \&
  S{\'a}ri}]{Noyes:2008p12477}
Noyes, R.~W., Bakos, G.~{\'A}., Torres, G., {et~al.} 2008, ApJ, 673, L79

\bibitem[{Nutzman {et~al.}(2011)Nutzman, Gilliland, McCullough, Charbonneau,
  Christensen-Dalsgaard, Kjeldsen, Nelan, Brown, \&
  Holman}]{Nutzman:2011p12545}
Nutzman, P., Gilliland, R.~L., McCullough, P.~R., {et~al.} 2011, ApJ, 726, 3

\bibitem[{Ogilvie \& Lin(2004)}]{Ogilvie:2004p11737}
Ogilvie, G.~I. \& Lin, D. N.~C. 2004, ApJ, 610, 477

\bibitem[{P{\'a}l {et~al.}(2010)P{\'a}l, Bakos, Torres, Noyes, Fischer,
  Johnson, Henry, Butler, Marcy, Howard, Sip{\H o}cz, Latham, \&
  Esquerdo}]{Pal:2010p12466}
P{\'a}l, A., Bakos, G.~{\'A}., Torres, G., {et~al.} 2010, MNRAS, 401, 2665

\bibitem[{P{\'a}l {et~al.}(2008)P{\'a}l, Bakos, Torres, Noyes, Latham,
  Kov{\'a}cs, Marcy, Fischer, Butler, Sasselov, Sip{\H o}cz, Esquerdo,
  Kov{\'a}cs, Stefanik, L{\'a}z{\'a}r, Papp, \& S{\'a}ri}]{Pal:2008p4127}
P{\'a}l, A., Bakos, G.~{\'A}., Torres, G., {et~al.} 2008, ApJ, 680, 1450

\bibitem[{Pollacco {et~al.}(2008)Pollacco, Skillen, Cameron, Loeillet,
  Stempels, Bouchy, Gibson, Hebb, H{\'e}brard, Joshi, McDonald, Smalley, Smith,
  Street, Udry, West, Wilson, Wheatley, Aigrain, Alsubai, Benn, Bruce,
  Christian, Clarkson, Enoch, Evans, Fitzsimmons, Haswell, Hellier, Hickey,
  Hodgkin, Horne, Hrudkov{\'a}, Irwin, Kane, Keenan, Lister, Maxted, Mayor,
  Moutou, Norton, Osborne, Parley, Pont, Queloz, Ryans, \&
  Simpson}]{Pollacco:2008p5241}
Pollacco, D., Skillen, I., Cameron, A.~C., {et~al.} 2008, MNRAS, 385, 1576

\bibitem[{Queloz {et~al.}(2010)Queloz, Anderson, Cameron, Gillon, Hebb,
  Hellier, Maxted, Pepe, Pollacco, S{\'e}gransan, Smalley, Triaud, Udry, \&
  West}]{Queloz:2010p7376}
Queloz, D., Anderson, D., Cameron, A.~C., {et~al.} 2010, A{\&}A, 517, L1

\bibitem[{Queloz {et~al.}(2000)Queloz, Eggenberger, Mayor, Perrier, Beuzit,
  Naef, Sivan, \& Udry}]{Queloz:2000p247}
Queloz, D., Eggenberger, A., Mayor, M., {et~al.} 2000, A{\&}A, 359, L13

\bibitem[{Rasio \& Ford(1996)}]{Rasio:1996p3680}
Rasio, F.~A. \& Ford, E.~B. 1996, Science, 274, 954

\bibitem[{Rossiter(1924)}]{Rossiter:1924p869}
Rossiter, R.~A. 1924, ApJ, 60, 15

\bibitem[{Sanchis-Ojeda \& Winn(2011)}]{SanchisOjeda:2011p12868}
Sanchis-Ojeda, R. \& Winn, J.~N. 2011, eprint arXiv, 1107, 2920, submitted to
  ApJ [9 pages]

\bibitem[{Sato {et~al.}(2005)Sato, Fischer, Henry, Laughlin, Butler, Marcy,
  Vogt, Bodenheimer, Ida, Toyota, Wolf, Valenti, Boyd, Johnson, Wright, Ammons,
  Robinson, Strader, McCarthy, Tah, \& Minniti}]{Sato:2005p2510}
Sato, B., Fischer, D.~A., Henry, G.~W., {et~al.} 2005, ApJ, 633, 465

\bibitem[{Schlaufman(2010)}]{Schlaufman:2010p7946}
Schlaufman, K.~C. 2010, ApJ, 719, 602

\bibitem[{Shporer {et~al.}(2009)Shporer, Bakos, Bouchy, Pont, Kov{\'a}cs,
  Latham, Sip{\"o}cz, Torres, Mazeh, Esquerdo, P{\'a}l, Noyes, Sasselov,
  L{\'a}z{\'a}r, Papp, S{\'a}ri, \& Kov{\'a}cs}]{Shporer:2009p12233}
Shporer, A., Bakos, G.~{\'A}., Bouchy, F., {et~al.} 2009, ApJ, 690, 1393

\bibitem[{Simpson {et~al.}(2011)Simpson, Pollacco, Cameron, H{\'e}brard,
  Anderson, Barros, Boisse, Bouchy, Faedi, Gillon, Hebb, Keenan, Miller,
  Moutou, Queloz, Skillen, Sorensen, Stempels, Triaud, Watson, \&
  Wilson}]{Simpson:2011p8894}
Simpson, E.~K., Pollacco, D., Cameron, A.~C., {et~al.} 2011, MNRAS, 600

\bibitem[{Southworth {et~al.}(2011)Southworth, Dominik, J{\o}rgensen, Rahvar,
  Snodgrass, Alsubai, Bozza, Browne, Burgdorf, Novati, Dodds, Dreizler, Finet,
  Gerner, Hardis, Harps{\o}e, Hellier, Hinse, Hundertmark, Kains, Kerins,
  Liebig, Mancini, Mathiasen, Penny, Proft, Ricci, Sahu, Scarpetta,
  Sch{\"a}fer, Sch{\"o}nebeck, \& Surdej}]{Southworth:2011p12650}
Southworth, J., Dominik, M., J{\o}rgensen, U.~G., {et~al.} 2011, A{\&}A, 527, 8

\bibitem[{Sozzetti {et~al.}(2007)Sozzetti, Torres, Charbonneau, Latham, Holman,
  Winn, Laird, \& O'Donovan}]{Sozzetti:2007p2647}
Sozzetti, A., Torres, G., Charbonneau, D., {et~al.} 2007, ApJ, 664, 1190

\bibitem[{Torres {et~al.}(2010)Torres, Bakos, Hartman, Kov{\'a}cs, Noyes,
  Latham, Fischer, Johnson, Marcy, Howard, Sasselov, Kipping, Sip{\H o}cz,
  Stefanik, Esquerdo, Everett, L{\'a}z{\'a}r, Papp, \&
  S{\'a}ri}]{Torres:2010p12511}
Torres, G., Bakos, G.~{\'A}., Hartman, J., {et~al.} 2010, ApJ, 715, 458

\bibitem[{Triaud {et~al.}(2010)Triaud, Cameron, Queloz, Anderson, Gillon, Hebb,
  Hellier, Loeillet, Maxted, Mayor, Pepe, Pollacco, S{\'e}gransan, Smalley,
  Udry, West, \& Wheatley}]{Triaud:2010p8039}
Triaud, A. H. M.~J., Cameron, A.~C., Queloz, D., {et~al.} 2010, A{\&}A, 524, 25

\bibitem[{Triaud {et~al.}(2009)Triaud, Queloz, Bouchy, Moutou, Cameron, Claret,
  Barge, Benz, Deleuil, Guillot, H{\'e}brard, {\'E}tangs, Lovis, Mayor, Pepe,
  \& Udry}]{Triaud:2009p7520}
Triaud, A. H. M.~J., Queloz, D., Bouchy, F., {et~al.} 2009, A{\&}A, 506, 377

\bibitem[{Ward(1997)}]{Ward:1997p11274}
Ward, W.~R. 1997, Icarus, 126, 261

\bibitem[{Watson {et~al.}(2011)Watson, Littlefair, Diamond, Cameron,
  Fitzsimmons, Simpson, Moulds, \& Pollacco}]{Watson:2011p11844}
Watson, C.~A., Littlefair, S.~P., Diamond, C., {et~al.} 2011, MNRAS, 413, L71

\bibitem[{Winn {et~al.}(2010{\natexlab{a}})Winn, Fabrycky, Albrecht, \&
  Johnson}]{Winn:2010p7311}
Winn, J.~N., Fabrycky, D., Albrecht, S., \& Johnson, J.~A. 2010{\natexlab{a}},
  ApJ, 718, L145

\bibitem[{Winn {et~al.}(2011)Winn, Howard, Johnson, Marcy, Isaacson, Shporer,
  Bakos, Hartman, Holman, Albrecht, Crepp, \& Morton}]{Winn:2011p11042}
Winn, J.~N., Howard, A.~W., Johnson, J.~A., {et~al.} 2011, AJ, 141, 63

\bibitem[{Winn {et~al.}(2009{\natexlab{a}})Winn, Johnson, Albrecht, Howard,
  Marcy, Crossfield, \& Holman}]{Winn:2009p3712}
Winn, J.~N., Johnson, J.~A., Albrecht, S., {et~al.} 2009{\natexlab{a}}, ApJ,
  703, L99

\bibitem[{Winn {et~al.}(2009{\natexlab{b}})Winn, Johnson, Fabrycky, Howard,
  Marcy, Narita, Crossfield, Suto, Turner, Esquerdo, \&
  Holman}]{Winn:2009p3777}
Winn, J.~N., Johnson, J.~A., Fabrycky, D., {et~al.} 2009{\natexlab{b}}, ApJ,
  700, 302

\bibitem[{Winn {et~al.}(2010{\natexlab{b}})Winn, Johnson, Howard, Marcy, Bakos,
  Hartman, Torres, Albrecht, \& Narita}]{Winn:2010p11036}
Winn, J.~N., Johnson, J.~A., Howard, A.~W., {et~al.} 2010{\natexlab{b}}, ApJ,
  718, 575

\bibitem[{Winn {et~al.}(2010{\natexlab{c}})Winn, Johnson, Howard, Marcy,
  Isaacson, Shporer, Bakos, Hartman, \& Albrecht}]{Winn:2010p8446}
Winn, J.~N., Johnson, J.~A., Howard, A.~W., {et~al.} 2010{\natexlab{c}}, ApJ,
  723, L223

\bibitem[{Wolf {et~al.}(2007)Wolf, Laughlin, Henry, Fischer, Marcy, Butler, \&
  Vogt}]{Wolf:2007p1867}
Wolf, A.~S., Laughlin, G., Henry, G.~W., {et~al.} 2007, ApJ, 667, 549

\bibitem[{Wu \& Lithwick(2011)}]{Wu:2011p14507}
Wu, Y. \& Lithwick, Y. 2011, ApJ, 735, 109

\bibitem[{Wu {et~al.}(2007)Wu, Murray, \& Ramsahai}]{Wu:2007p4179}
Wu, Y., Murray, N.~W., \& Ramsahai, J.~M. 2007, ApJ, 670, 820

\bibitem[{Zahn(1977)}]{Zahn:1977p14439}
Zahn, J.-P. 1977, A{\&}A, 57, 383

\end{thebibliography}
\end{document}